\documentclass[notitlepage,aps,pra,onecolumn,superscriptaddress,amsmath,showpacs,tightenlines,longbibliography,11pt]{revtex4-1}
\usepackage{amssymb}
\usepackage{amsmath}
\usepackage{dcolumn}
\usepackage{graphicx}
\usepackage{mathrsfs}
\usepackage{subfigure}
\usepackage{booktabs}
\usepackage{color}
\usepackage{docmute}
\usepackage{hyphenat}
\usepackage{CJKutf8}
\usepackage{multirow}
\usepackage{enumerate}

\newdimen\origiwspc
\newdimen\origiwstr

\newcommand{\ket}[1]{\mbox{$|#1\rangle$}}
\newcommand{\bra}[1]{\mbox{$\langle#1|$}}
\newcommand{\average}[1]{\mbox{$\langle#1\rangle$}}

\makeatletter
\newcommand*\bigcdot{\mathpalette\bigcdot@{.5}}
\newcommand*\bigcdot@[2]{\mathbin{\vcenter{\hbox{\scalebox{#2}{$\m@th#1\bullet$}}}}}
\makeatother

\makeatletter
\newcommand{\raisemath}[1]{\mathpalette{\raisem@th{#1}}}
\newcommand{\raisem@th}[3]{\raisebox{#1}{$#2#3$}}
\makeatother

\usepackage{url}
\usepackage[colorlinks]{hyperref}
\hypersetup{%
	plainpages=true,
	breaklinks=true,
	hypertexnames=false,
	pageanchor=true,
	bookmarksnumbered=true,
	colorlinks=true,
	linkcolor={blue},
	citecolor={red},
	urlcolor={blue},
	anchorcolor={black}
}

\begin{document}
	

\title{Proposal of ensemble qubits with two-atom decay}

\author{Wei Qin}
\email{qinwei09@tsinghua.org.cn}
\affiliation{Theoretical Quantum Physics Laboratory, Cluster
	for Pioneering Research, RIKEN, Wako-shi, Saitama 351-0198, Japan}
\affiliation{Center for Joint Quantum Studies and Department of Physics, School of Science, Tianjin University, Tianjin 300350, China}

\author{Adam Miranowicz}
\affiliation{Theoretical Quantum Physics Laboratory, Cluster
	for Pioneering Research, RIKEN, Wako-shi, Saitama 351-0198, Japan}
\affiliation{Institute of Spintronics and Quantum Information, Faculty of Physics, Adam Mickiewicz University, 61-614 Pozna\'n, Poland}

\author{Franco Nori}
\affiliation{Theoretical Quantum Physics Laboratory, Cluster
	for Pioneering Research, RIKEN, Wako-shi, Saitama 351-0198, Japan}
\affiliation{Center for Quantum Computing, RIKEN, Wako-shi, Saitama 351-0198, Japan}
\affiliation{Department of Physics, The University of Michigan,	Ann Arbor, Michigan 48109-1040, USA}

\begin{abstract}
We propose and analyze a novel approach to implement ensemble qubits. The required anharmonicity is provided by a simultaneous decay of two atoms (i.e., two-atom decay), which is achieved by fully quantum degenerate parametric amplification. For an atomic ensemble, the two-atom decay generates and stabilizes a 2D quantum manifold, which is spanned by the ground and single-excited superradiant states. Moreover, this nonlinear decay process can strongly suppress transitions to higher-excited superradiant states, and convert residual transitions into an effective decay from the single-excitation superradiant state to the ground state. Our method does not require Rydberg dipole blockade and, thus, strong atom-atom interactions, compared to previous work. This indicates that it can be applied to typical atomic or spin ensembles in simple experimental setups. Remarkably, our idea is compatible with the cavity protection mechanism, and therefore spin dephasing due to inhomogeneous broadening can be strongly suppressed. The presented ensemble qubit provides a new platform for quantum information processing, and also extends the range of applications of atomic or spin ensembles.
\end{abstract}

\date{\today}

\maketitle

\section{Introduction}
Ensembles of atoms or spins have been attracting considerable
interest, because of their long coherence times (even at room temperatures) and their ability to interface between the microwave and optical regimes. Moreover, ensembles of atoms or spins have been used to, e.g., build quantum repeaters for long distance quantum communication~\cite{duan2001long,kuzmich2003generation,hammerer2010quantum,sangouard2011quantum}, design hybrid systems for powerful quantum computation~\cite{xiang2013hybrid,kurizki2015quantum,gu2017microwave,clerk2020hybrid}, and generate nonclassical states for high-precision quantum measurements~\cite{kitagawa1993squeezed,agarwal1997atomic,hald1999spin,ma2011quantum,pezze2018quantum,qin2020strong,groszkowski2020heisenberg,qin2021generating,liu2022generation}. Despite such considerable
developments, exploiting an ensemble as a single qubit remains a challenge, due to the requirement of a strong anharmonicity. Previous approaches, which have been proposed to encode an ensemble qubit, usually rely on the use of Rydberg atoms~\cite{dipole2001lukin,brion2007quantum,moller2008quantum,yan2008quantum,mei2009scalable,zwierz2010applications,saffman2010quantum,ebert2015coherence,saffman2016quantum,spong2021collectively}. In these approaches, the strong anharmonicity is provided by Rydberg blockade based on strong dipole-dipole interactions, which require atoms to be excited to their highly excited states using multiphoton processes. For example, alkali atoms need to be excited to their states with the principal quantum number $n>70$ to achieve Rydberg dipole blockade. However, for typical ensembles, it is much easier to control atoms in weakly excited states, rather than in highly excited states. Furthermore, the interactions between atoms in  weakly excited states are extremely weak, such that these atoms are often considered to be effectively independent. Thus, a new mechanism, which can generate a strong anharmonicity in ensembles of atoms in weakly excited states, is desirable for the formation of ensemble qubits.  

To address the above issue, we describe in this manuscript the use of the two-atom decay to provide a strong anharmonicity. The two-atom decay, proposed in our recent work~\cite{qin2021generating} by exploiting a fully quantum degenerate parametric amplifier (DPA), refers to a simultaneous decay of two atoms in an ensemble. As usual, we assume the atomic ensemble to be in the low-excitation regime, where the number of excited atoms is much smaller than the total number of atoms. In such a regime, the spin-wave approximation can be applied, and the ensemble behaves as a harmonic oscillator. However, the nonlinear two-atom decay conserves the parity of the number of the excited atoms, and as a result the ensemble states are restricted to a 2D quantum manifold, spanned by the ground state and a symmetric collective state with only one excitation (i.e., a single-excitation superradiant state). Moreover, due to the two-atom decay, the transitions out of the 2D quantum subspace (e.g., driven by an external driving) are strongly suppressed, and the residual transitions are converted into an effective decay from the single-excitation superradiant state to the ground state. This indicates that an ensemble qubit is formed. Such an ensemble qubit is not discussed in our previous proposal in Ref.~\cite{qin2021generating}, although a 2D quantum manifold spanned by long-lived atomic cat states has been shown there. Note that compared to Ref.~\cite{qin2021generating}, our present proposal, though as an extension of Ref.~\cite{qin2021generating}, involves no pump-mode driving, as a result no two-atom excitation, and thus cannot stabilize atomic cat states.

We also show that our proposal is compatible with the cavity protection technique, such that the resulting ensemble qubit can be well protected from inhomogeneous broadening, which is the main source of noise of atomic ensembles. Specifically, a strong coupling between the ensemble and the signal mode makes the superradiant spin-wave mode largely detuned and, thus, decoupled from the subradiant spin-wave modes. Consequently, spin dephasing due to inhomogeneous broadening can be strongly suppressed. In contrast to previous work on ensemble qubits, our proposal requires neither Rydberg dipole blockade nor strong atom-atom interactions. Moreover, multiphoton processes used to excite and control Rydberg atoms are not required either. Therefore, it could be applied to common atomic or spin ensembles, thus enabling a significant simplification of the experimental complexity. 

The remaining of the paper is organized as follows. The physical system used to form an ensemble qubit, i.e., an atomic ensemble coupled to the signal mode of a fully quantum DPA, is introduced in Sec.~\ref{physical model}. Included are the system Hamiltonian and the corresponding master equation. In Sec.~\ref{Ensemble qubit}, we discuss a parametric coupling between an atom pair and a pump-mode photon, obtain the two-atom decay by adiabatically eliminating the pump mode, and then, with such a nonlinear atomic decay, analyze the form of an ensemble qubit. Section~\ref{Rabi oscillations} presents Rabi oscillations of the ensemble qubit subject to an external field, and shows that the transitions out of the ensemble-qubit subspace, driven by this external field, can be converted to an effective decay process of the ensemble qubit. We demonstrate in Sec.~\ref{Inhomogeneous broadening} that our idea is compatible with the cavity protection mechanism, and thus can strongly suppress spin dephasing due to inhomogeneous broadening for the ensemble qubit. Conclusions are drawn in Sec.~\ref{Conclusions}. Finally, three Appendixes provide some detailed derivations.

\begin{figure}[b]
	\centering
	\includegraphics[width=8.0cm]{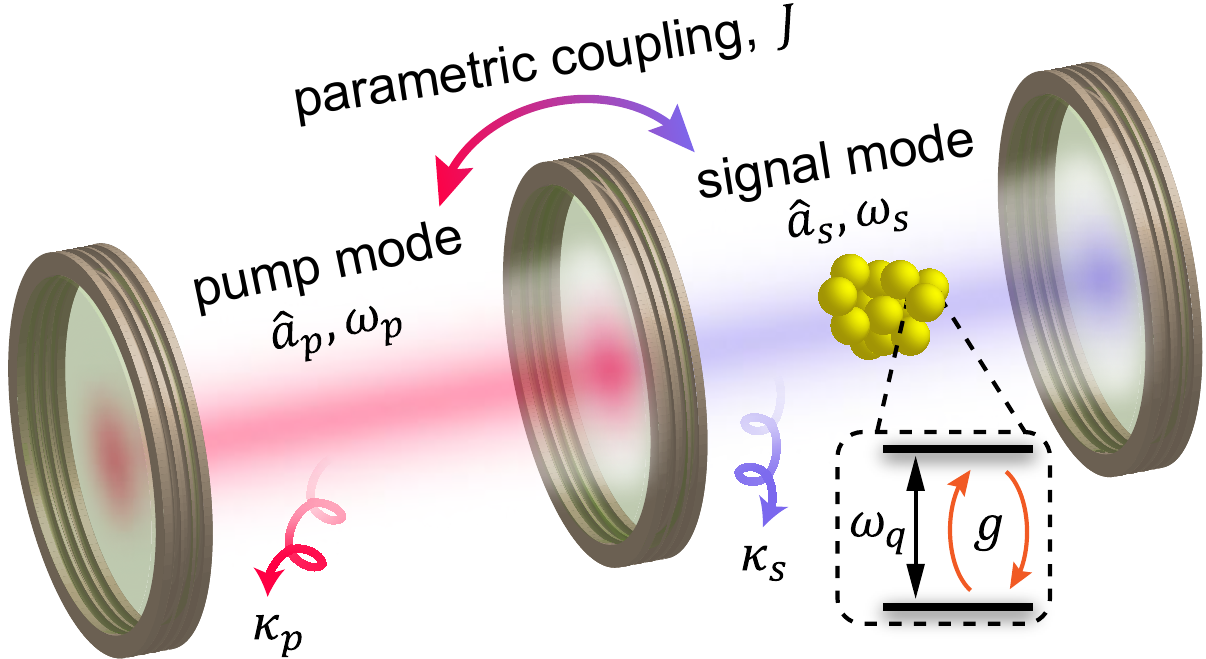}
	\caption{Schematics of our setup used to form an ensemble qubit with a fully quantum degenerate parametric amplifier. The pump and signal modes, $\hat{a}_{p}$ and $\hat{a}_{s}$, coupled via a single-photon parametric coupling of strength $J$, are represented by two cavities, for simplicity. An ensemble consists of $N$ two-level atoms, each with a transition frequency $\omega_{q}$, and it is coupled to the signal mode with a single-atom coupling of strength $g$. Here, $\omega_{p}$, $\omega_{s}$ are the resonance frequencies of the pump and signal modes, and $\kappa_{p}$, $\kappa_{s}$ are their respective photon loss rates.}\label{fig_schematics}
\end{figure}

\section{physical model}
\label{physical model}
The basic idea is schematically illustrated in Fig.~\ref{fig_schematics}. We consider a fully quantum DPA, with a pump mode $\hat{a}_{p}$ of frequency $\omega_{p}$, and a signal mode $\hat{a}_{s}$ of frequency $\omega_{s}$. These two modes are coupled through a single-photon parametric coupling of strength $J$. We further assume that an ensemble of $N$ two-level atoms is coupled to the signal mode with a single-atom strength $g$, and that all the atoms have the same transition frequency $\omega_{q}$. The system can thus be described by the Hamiltonian,
\begin{align}\label{eq_full_Hamiltonian}
\hat{H}=\omega_{s}\hat{a}_{s}^{\dagger}\hat{a}_{s}+\omega_{p}\hat{a}_{p}^{\dagger}\hat{a}_{p}+\omega_{q}\hat{S}_{z}+g\left(\hat{a}_{s}^{\dagger}\hat{S}_{-}+\hat{a}_{s}\hat{S}_{+}\right)+J\left(\hat{a}_{p}\hat{a}_{s}^{\dagger 2}+\hat{a}_{p}^{\dagger}\hat{a}_{s}^{2}\right),
\end{align}
where $\hat{S}_{z}=\frac{1}{2}\sum_{j=1}^{N}\sigma_{j}^{z}$ and $\hat{S}_{\pm}=\sum_{j=1}^{N}\sigma_{j}^{\pm}$ are collective spin operators of the ensemble. Here, $\sigma_{j}^{z}$ and $\sigma_{j}^{\pm}$ are the Pauli matrices of the $j$th atom. We have assumed, for simplicity, that the couplings of the atoms to the signal mode are uniform, but in Appendix~\ref{Collective spin operators for nonuniform single-atom couplings}, we demonstrate the collective spin operators also in the nonuniform case. Upon introducing the Holstein-Primakoff transformation~\cite{holstein1940field}, the collective spin can be mapped to a bosonic mode $\hat{s}$, i.e.,
\begin{align}
\hat{S}_{-}=\;&\hat{s}\sqrt{N-\hat{s}^{\dagger}\hat{s}},\\
\hat{S}_{z}=\;&-\frac{N}{2}+\hat{s}^{\dagger}\hat{s}.
\end{align}
We restrict our discussions to the case of $N\rightarrow\infty$, so that the number of excited atoms is much smaller than the total number of atoms, i.e., $\average{\hat{s}^{\dagger}\hat{s}}\ll N$. In this case, we can apply the spin-wave approximation and, as a consequence, have 
\begin{equation}\label{eq_superradiant_spin_wave_mode}
\hat{S}_{-}\simeq\sqrt{N}\hat{s}.
\end{equation}
The Hamiltonian $\hat{H}$ is accordingly transformed to 
\begin{align}\label{eq_full_Hamiltonian_boson}
\hat{\mathcal{H}}=\omega_{s}\hat{a}_{s}^{\dagger}\hat{a}_{s}+\omega_{p}\hat{a}_{p}^{\dagger}\hat{a}_{p}+\omega_{q}\hat{s}^{\dagger}\hat{s}+g_{\rm col}\left(\hat{a}_{s}^{\dagger}\hat{s}+\hat{a}_{s}\hat{s}^{\dagger}\right)+J\left(\hat{a}_{p}\hat{a}_{s}^{\dagger 2}+\hat{a}_{p}^{\dagger}\hat{a}_{s}^{2}\right),
\end{align}
where $g_{\rm col}=\sqrt{N}g$ is the strength of the collective coupling of the ensemble to the signal mode. Furthermore, we use the Lindblad dissipator $\mathcal{L}\left(o\right)\rho=\left(2\hat{o}\hat{\rho}\hat{o}^{\dagger}-\hat{o}^{\dagger}\hat{o}\hat{\rho}-\hat{\rho}\hat{o}^{\dagger}\hat{o}\right)/2$ to describe photon losses. Therefore, the full dynamics of the system can be determined by the following master equation:
\begin{equation}\label{eq_full_master_equation}
\dot{\hat{\rho}}=-i\left[\hat{\mathcal{H}},\hat{\rho}\right]+\kappa_{p}\mathcal{L}\left(\hat{a}_{p}\right)\hat{\rho}+\kappa_{s}\mathcal{L}\left(\hat{a}_{s}\right)\hat{\rho},
\end{equation}
where $\kappa_{p}$ and $\kappa_{s}$ are the photon loss rates of the pump and signal modes, respectively. 

Note that for typical ensembles, the single-atom coupling $g$ is extremely weak, and thus an extremely large $N$ is required to induce a strong $g_{\rm col}$. For example, for spin ensembles of nitrogen-vacancy (NV) centers, having $g\simeq2\pi\times10$~Hz and $N\simeq10^{12}$ results in $g_{\rm col}\simeq2\pi\times10$~MHz. In this case, the specific value of $N$ is not easy to determine, and is not so important. In fact, the key element is the collective coupling $g_{\rm col}$, which can be well measured in experiments and is usually treated as a constant for a given setup. Thus, here we do not have to take into account the dependence of our proposal on the number $N$ of atoms. Instead, finding such a dependence, as a research agenda in the future, may be more important if our proposal is used for small-size ensembles (e.g., $N\sim10$ or $100$) with a strong coupling $g$.

Furthermore, in the general case, the single-photon coupling $J$ is also very weak, so that a strong pump-mode driving is needed to generate a strong parametric (i.e., two-photon) driving for the signal mode. In this case, the pump mode is treated classically and its quantum nature is neglected~\cite{scully1997book,drummond2004quantum,drummond2014the}. Such a type of DPA is called semiclassical, and its intracavity squeezing has been extensively studied for various applications~\cite{huang2009enhancement, lu2015squeezed,qin2018exponentially,leroux2018enhancing,ge2019trapped,qin2019proposal,qin2019emission,chen2019fast,li2020enhancing,burd2021quantum,tang2022quantum,peano2015intracavity,zhou2021n,chen2021shortcuts,chen2022fault}. For example, applying a
semiclassical DPA to the Jaynes-Cummings model can enhance the ability of cavity and circuit quantum electrodynamics systems to process quantum information, which was predicted
in Refs.~\cite{qin2018exponentially,leroux2018enhancing,ge2019trapped} and demonstrated
experimentally in Ref.~\cite{burd2021quantum}. Recently, with the rapid development of quantum devices, a strong coupling $J$, which can range from several tens of kHz to several tens of MHz~\cite{leghtas2015confining,touzard2018coherent,lescanne2020exponential,chang2020observation,vrajitoarea2020quantum,guo2016second,bruch2019chip,wang2021efficient}, has been achieved in experiments. This strong single-photon coupling allows to exploit the quantum nature of the pump mode, i.e., a fully quantum DPA, for modern quantum technologies~\cite{gerry1993generation, gilles1994generation,mirrahimi2014dynamically,guillaud2019repetition,puri2019stabilized,sun2019schrodinger,qin2022beating,cai2021bosonic,ma2021quantum,zhou2022enhancing}. As demonstrated below, a fully quantum DPA can be used to confine the state of an atomic ensemble to a 2D quantum manifold spanned by the ground state and a single-excitation superradiant state, forming an ensemble qubit.

\section{Ensemble qubit}
\label{Ensemble qubit}
\begin{figure}[t]
	\centering
	\includegraphics[width=16cm]{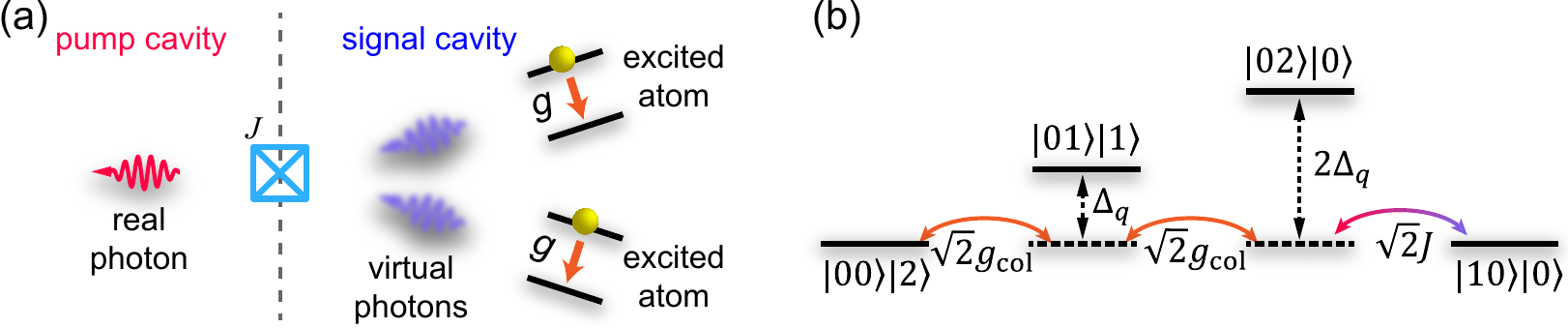}
	\caption{(a) Parametric coupling between an atom pair and a real pump photon mediated by a virtual signal-photon pair. Two excited atoms emit two virtual signal photons, which then are up-converted into a real pump photon. (b) Transition processes corresponding to the two-atom single-photon conversion shown in (a).}\label{fig_transitions}
\end{figure}

In this section, we show how to form a two-level ensemble, i.e., an ensemble qubit, by using a fully quantum DPA. To proceed, we assume that both couplings of the signal mode to the ensemble and to the pump mode are largely detuned, i.e.,  
\begin{align}
\omega_{s}-\omega_{q}\equiv\;&\Delta_{q}\gg g_{\rm col}, \\ 2\omega_{s}-\omega_{p}\equiv\;&\Delta_{p}\gg J.
\end{align}
But at the same time, the pump mode frequency is approximately equal to the doubled atomic transition frequency, i.e., 
\begin{equation}
\omega_{p}\simeq 2\omega_{q}.
\end{equation}
Note that here, a slight detuning between $\omega_{p}$ and $2\omega_{q}$ has been assumed to compensate some undesired resonance shifts of the atoms and the pump mode (see below for details).
Under these assumptions, we can predict a parametric coupling between atom pairs and real pump photons, i.e., the joint absorption and emission of a pump photon by two atoms~\cite{garziano2016one,garziano2020atoms}. This parametric coupling is mediated by a virtual signal-photon pair, as shown in Fig.~\ref{fig_transitions}(a). Note that the virtual-photon mediated coupling has been previously exploited for quantum nonlinear optics~\cite{kockum2019ultrastrong,kockum2017deterministic,stassi2017quantum}. 

To better understand the form of the above parametric coupling, we demonstrate the virtual intermediate transitions between the states $\ket{00}\ket{2}$ and $\ket{10}\ket{0}$ in Fig.~\ref{fig_transitions}(b). Here, the first ket $\ket{m_{p}m_{s}}$ ($m_{p}, m_{s}=0,1,2,\cdots$) in the pair accounts for the DPA state with $m_{p}$ photons in the pump mode and $m_{s}$ photons in the signal mode, and the second ket $\ket{n}$ ($n=0,1,2,\cdots$) accounts for the symmetric superradiant state, $\ket{S=N/2,m_{z}=-N/2+n}$, of the collective spin, with $n$ excited atoms. Note that such a collective superradiant state can be formed as long as $N\geq2$. The coupling $g$ drives two virtual transitions in turn from the state $\ket{00}\ket{2}$, via the state $\ket{01}\ket{1}$ with a detuning $\Delta_{q}=\omega_{s}-\omega_{q}$, to the state $\ket{02}\ket{0}$ detuned by $2\Delta_{q}$. Subsequently, a virtual transition from $\ket{02}\ket{0}$ to the real state $\ket{10}\ket{0}$ is driven by the parametric coupling $J$. 

\begin{figure}[t]
	\centering
	\includegraphics[width=16cm]{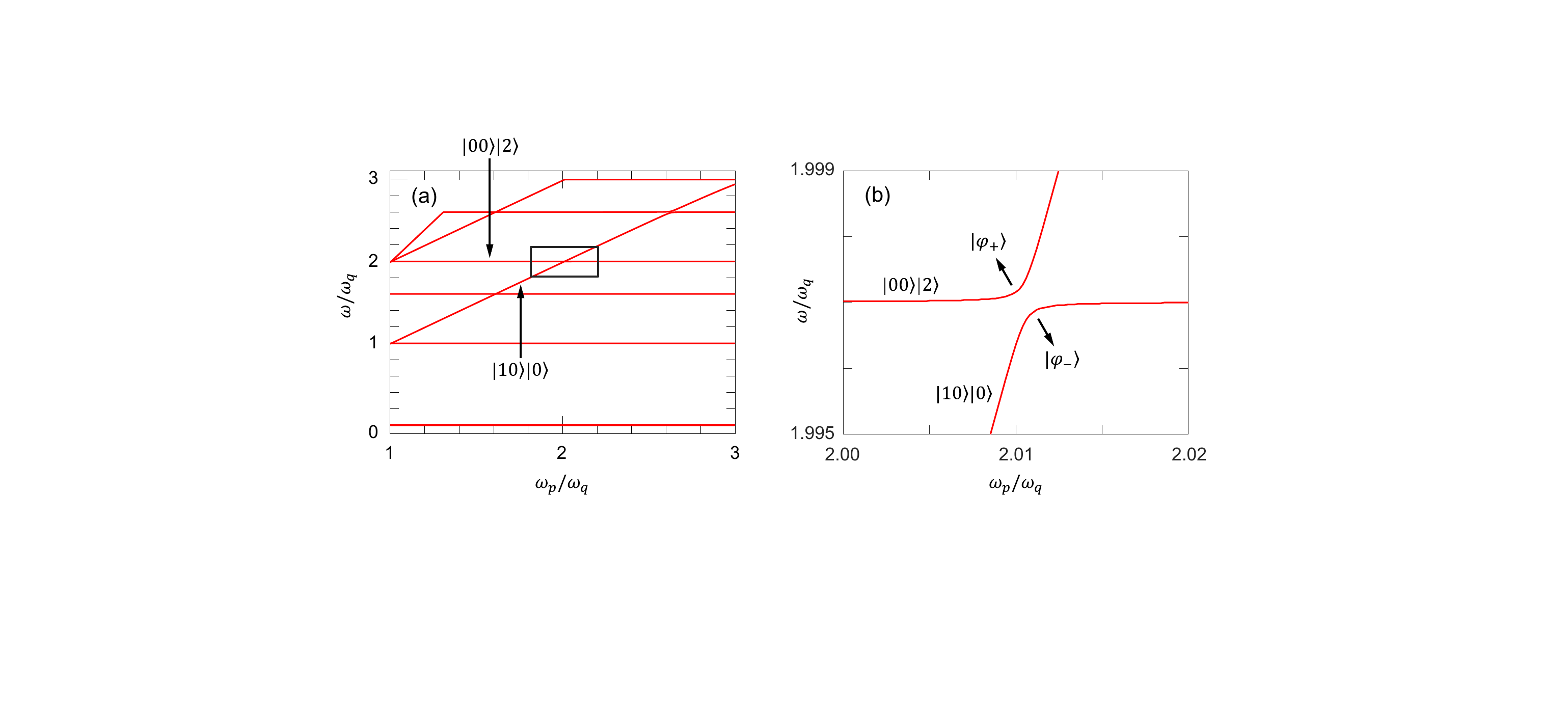}
	\caption{(a) Lowest energy levels of the system Hamiltonian in Eq.~(\ref{eq_full_Hamiltonian_boson}) as a function of $\omega_{p}/\omega_{q}$. The two levels indicated by the arrows corresponds to the states $\ket{10}\ket{0}$ and $\ket{00}\ket{2}$, respectively. Here, we set $g_{\rm col}=0.03\omega_{q}$, $J=3g_{\rm col}$, and $\Delta_{q}=20g_{\rm col}$. (b)~Enlarged view of the boxed region in (c), showing an avoided level crossing around $\omega_{p}/\omega_{q}=2$. This avoided crossing originates from the hybridization of the states $\ket{00}\ket{2}$ and $\ket{10}\ket{0}$ due to the parametric coupling $\chi$ in Eq.~(\ref{H_avg}).}\label{fig_splitting}
\end{figure}

Thus, the Hamiltonian $\mathcal{H}$ in Eq.~(\ref{eq_full_Hamiltonian_boson}), after time averaging in the interaction picture (see Appendix~\ref{Derivation of the effective Hamiltonian} for details), becomes
\begin{equation}\label{H_avg}
\hat{\mathcal{H}}_{\rm avg}=\chi\left(\hat{a}_{p}\hat{s}^{\dagger 2}+\hat{a}_{p}^{\dagger}\hat{s}^{2}\right),
\end{equation}
where  
\begin{equation}
\chi=\frac{g_{\rm col}^{2}J}{\Delta_{q}^{2}}
\end{equation}
is the strength of an effective parametric coupling between the ensemble and the pump mode. We numerically diagonalize the system Hamiltonian $\hat{\mathcal{H}}$ in Eq.~(\ref{eq_full_Hamiltonian_boson}), and plot the lowest energy levels in Fig.~\ref{fig_splitting}(a). It seems as though the levels of the states $\ket{10}\ket{0}$ and $\ket{00}\ket{2}$ cross around $\omega_{p}/\omega_{q}=2$. However, in fact, the predicted parametric coupling $\chi$ can hybridize the states $\ket{10}\ket{0}$ and $\ket{00}\ket{2}$ and, as a result, induce an avoided crossing, with two hybridized states 
\begin{equation}
\ket{\varphi_{\pm}}=\frac{1}{\sqrt{2}}\left(\ket{10}\ket{0}\pm\ket{00}\ket{2}\right).
\end{equation}
This avoided level crossing becomes apparent in the enlarged view shown in Fig.~\ref{fig_splitting}(b). 

Let us now consider how to implement a two-level atomic ensemble with the time-averaged Hamiltonian $\hat{\mathcal{H}}_{\rm avg}$. We assume that the photon loss of the pump mode is very strong, so that the pump photon emitted jointly by two excited atoms is lost rather than reabsorbed by these two atoms. This process leads to the desired two-atom decay. To be more explicit, we adiabatically eliminate the pump mode $\hat{a}_{p}$ of the time-averaged Hamiltonian $\hat{\mathcal{H}}_{\rm avg}$ (see Appendix~\ref{Adiabatic elimination of the pump mode}), and then obtain an adiabatic master equation associated only with the degree of freedom of the ensemble, i.e.,
\begin{equation}\label{eq_two_atom_decay}
\dot{\hat{\rho}}_{\rm ens}=\kappa_{\rm 2at}\mathcal{L}\left(\hat{s}^{2}\right)\hat{\rho}_{\rm ens},
\end{equation}
where
\begin{equation}
\kappa_{\rm 2at}=\frac{4\chi^{2}}{\kappa_{p}}
\end{equation}
is the rate of the two-atom decay. It is clear that the two-atom decay conserves the parity of the number of excited atoms. Therefore, the state space of the ensemble can be decomposed into even and odd subspaces, according to the eigenvalues $\pm1$ of the parity operator $P=\exp\left(i\pi\hat{s}^{\dagger}\hat{s}\right)$. Under the master equation in Eq.~(\ref{eq_two_atom_decay}), the ensemble can be cooled to its ground state $\ket{0}$ for an even-parity initial state, and to a single-excitation superradiant state $\ket{1}$ for an odd-parity initial state. Here, the ket $\ket{n}$ ($n=0,1,2,\dots$) refers to the collective spin state $\ket{S=N/2,m_{z}=-N/2+n}$ as already mentioned above, such that
\begin{equation}
\ket{n}\equiv\ket{S=N/2,m_{z}=-N/2+n}.
\end{equation}
This indicates that the steady state of the ensemble is restricted by the two-atom decay into a 2D quantum manifold $\mathcal{C}$ spanned by $\ket{0}$ and $\ket{1}$, and can be expressed as 
\begin{align}
\hat{\rho}_{\rm ens}^{\rm ss}=c_{00}\ket{0}\bra{0}+c_{11}\ket{1}\bra{1}+c_{01}\ket{0}\bra{1}+c_{10}\ket{1}\bra{0},
\end{align}
where $c_{i,j}$ ($i,j=0,1$) are some coefficients, which are determined below.

In order to determine the coefficients $c_{ij}$, we define the following two conserved quantities~\cite{simaan1978off,albert2014symmetries}:
\begin{align}
\hat{\Pi}_{00}=\;&\sum_{n=0}\ket{2n}\bra{2n},\\
\hat{\Pi}_{01}=\;&\frac{\left(\hat{s}^{\dagger}\hat{s}-1\right)!!}{\hat{s}^{\dagger}\hat{s}!!}\hat{\Pi}_{00}\hat{s},
\end{align}
where $n!!=n\times \left(n-2 \right)!!$ is the double factorial. Under the master equation in Eq.~(\ref{eq_two_atom_decay}), the time evolution of an operator $\hat{\Pi}$ in the Heisenberg picture is given by
\begin{equation}
\dot{\hat{\Pi}}=\frac{1}{2}\kappa_{\rm 2at}\left(2\hat{s}^{\dagger 2}\hat{\Pi}\hat{s}^{2}-\hat{s}^{\dagger 2}\hat{s}^{2}\hat{\Pi}-\hat{\Pi}\hat{s}^{\dagger 2}\hat{s}^{2}\right).
\end{equation}
After a straightforward calculation, we obtain
\begin{equation}
\dot{\hat{\Pi}}_{00}=\dot{\hat{\Pi}}_{01}=0,
\end{equation}
implying that these two quantities are indeed conserved.

According to Refs.~\cite{simaan1978off,albert2014symmetries}, the operators $\hat{\Pi}_{00}$ and $\hat{\Pi}_{01}$ determine the population, $c_{00}$, of the ground state $\ket{0}$, and the complex coherence, $c_{01}$, between $\ket{0}$ and the single-excitation superradiant state $\ket{1}$. As a result, the coefficients $c_{ij}$ can be calculated as
\begin{align}
\label{eq_c00}
c_{00}=\;&{\rm Tr}\left[\hat{\Pi}_{00}\hat{\rho}_{\rm ens}\left(0\right)\right],\\
\label{eq_c11}
c_{11}=\;&1-c_{00},\\
\label{eq_c01}
c_{01}=\;&{\rm Tr}\left[\hat{\Pi}_{01}^{\dagger}\hat{\rho}_{\rm ens}\left(0\right)\right],\\
\label{eq_c10}
c_{10}=\;&c_{01}^{*},
\end{align}
where $\hat{\rho}_{\rm ens}\left(0\right)$ is the initial state of the ensemble. Clearly, if the ensemble is initially in the even (or odd) parity subspace, we have $c_{00}=1$ (or $c_{11}=1$), and all other coefficients vanish.

\begin{figure}[b]
	\centering
	\includegraphics[width=8.3cm]{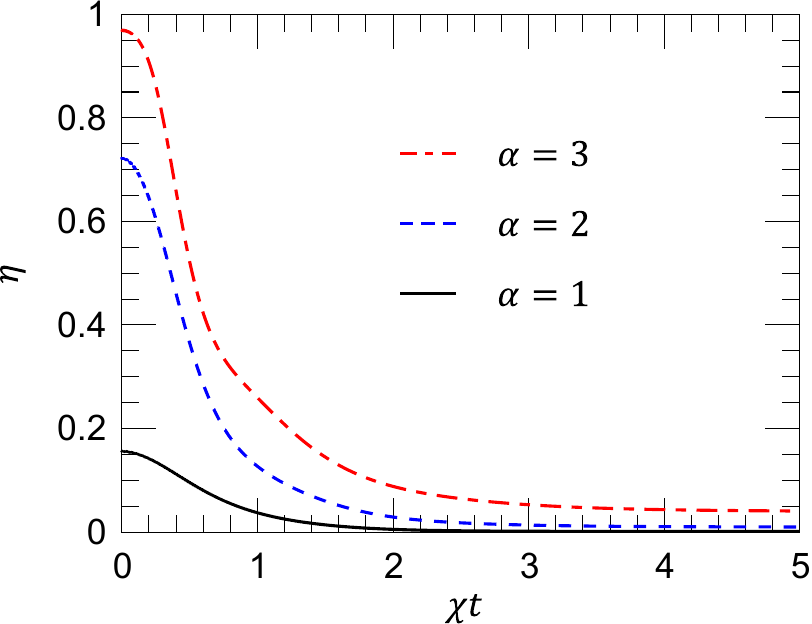}
	\caption{Time evolution of the state error $\eta$, with the ensemble being initially in different spin coherent states of coherent amplitudes $\alpha=1$, $2$, and $3$. We assumed that $J=3g_{\rm col}$, $\Delta_{q}=20g_{\rm col}$, $\kappa_{p}=5\chi$, $\kappa_{s}=0.3\kappa_{p}$, and that both the pump and signal modes are initialized in the vacuum.}\label{fig_fidelity_2D_manifold}
\end{figure}

In order to further confirm the steady-state prediction given above, we perform numerical simulations in Fig.~\ref{fig_fidelity_2D_manifold}. Specifically, we plot the time evolution of the state error $\eta=1-\mathcal{F}$, where $\mathcal{F}$ is the state fidelity. Initially, the ensemble is assumed to be in a spin coherent state $\hat{\rho}_{\rm ens}\left(0\right)=\ket{\theta,\phi}$, which is defined as 
\begin{equation}
\ket{\theta,\phi}=\hat{R}\left(\theta,\phi\right)\ket{0}.
\end{equation}
Here, 
\begin{equation}
\hat{R}\left(\theta,\phi\right)=\exp\left[\frac{1}{\sqrt{N}}\left(\alpha \hat{S}_{+}-\alpha^{*}\hat{S}_{-}\right)\right],
\end{equation}
where $\alpha=\sqrt{N}\theta\exp\left(i\phi\right)/2$, is a rotation operator of the collective spin, which rotates the ground state $\ket{0}$ of the ensemble by an angle $\theta$ about the axis $\left(\sin\phi, -\cos\phi,0\right)$ of the collective Bloch sphere. Clearly, under the spin-wave approximation (i.e., $\hat{S}_{-}\simeq\sqrt{N}\hat{s}$), the rotation operator $\hat{R}\left(\theta,\phi\right)$ becomes a bosonic displacement operator with a coherent amplitude $\alpha$, and accordingly, the spin coherent state $\ket{\theta, \phi}$ becomes approximated by a bosonic coherent state $\ket{\alpha}$. 
With such an initial state, a straightforward calculation gives
\begin{align}
\label{eq_coo_spin_coherent}
c_{00}=\;&\frac{1}{2}\left(1+e^{-2\left|\alpha\right|^{2}}\right),\\
\label{eq_co1_spin_coherent}
c_{01}=\;&\alpha^{*}e^{-\left|\alpha\right|^{2}}I_{0}\left(\left|\alpha\right|^{2}\right),
\end{align}
where $I_{0}\left(\bigcdot\right)$ is the modified Bessel function of the first kind.

In Fig.~\ref{fig_fidelity_2D_manifold}, we numerically integrate the original master equation in Eq.~(\ref{eq_full_master_equation}) for $\alpha=1$, $2$ and $3$ by using the QutiP toolbox~\cite{johansson2012qutip,johansson2013qutip2}, and obtain the actual states of the system. Then, we calculate the fidelity $\mathcal{F}$ between these actual states and the ideal states analytically predicted by the coefficients in Eq.~(\ref{eq_coo_spin_coherent}) and~(\ref{eq_co1_spin_coherent}). We find from Fig.~\ref{fig_fidelity_2D_manifold} that for realistic parameters of $g_{\rm col}/2\pi=10$~MHz and $J/2\pi=30$~MHz, an evolution time of $t=2/\chi\simeq0.42$~$\mu$s can keep the state error $\eta$ below $0.1$ even for an initial spin coherent state of $\alpha=3$. Thus, as expected, the ensemble states are driven by the two-atom decay into a stabilized 2D quantum manifold $\mathcal{C}$ of the states $\ket{0}$ and $\ket{1}$, with a high fidelity.

\section{Rabi oscillations}
\label{Rabi oscillations}
Having demonstrated the formation of a collective qubit, let us now discuss how to manipulate this qubit. We assume that the ensemble is driven by an external field of amplitude $\Omega_{d}$, phase $\theta_{d}$, and frequency $\omega_{d}$. As shown in Eq.~(\ref{eq_appendix_resonance_shift}) in Appendix~\ref{Derivation of the effective Hamiltonian}, the atomic transition frequency is shifted by an amount $\delta_{q}=g_{\rm col}^{2}/\Delta_{q}$, due to the second-order perturbation of the largely detuned Hamiltonian $\hat{\mathcal{H}}$ in Eq.~(\ref{eq_full_Hamiltonian_boson}). Thus, in order to make the ensemble driven resonantly by the external field, we set
\begin{equation}
\omega_{d}=\omega_{q}+\delta_{q}.
\end{equation}
This yields a driving Hamiltonian under the spin-wave approximation, i.e., 
\begin{equation}\label{eq_driving_Hamiltonian}
\hat{\mathcal{H}}_{d}=\Omega_{d}\left(e^{-i\theta_{d}}\hat{s}+e^{i\theta_{d}}\hat{s}^{\dagger}\right).
\end{equation}
Note that, here, $\hat{\mathcal{H}}_{d}$ has been transformed to the rotating frame, which the time-averaged Hamiltonian $\hat{\mathcal{H}}_{\rm avg}$ in Eq.~(\ref{H_avg}) and the adiabatic master equation in Eq.~(\ref{eq_two_atom_decay}) are in. 
Including the driving Hamiltonian $\hat{\mathcal{H}}_{d}$, the master equation in Eq.~(\ref{eq_two_atom_decay}) can be rewritten as,
\begin{equation}\label{eq_driving_two_atom_decay_master_equation}
\dot{\hat{\rho}}_{\rm ens}=-i\left[\hat{\mathcal{H}}_{d},\hat{\rho}_{\rm ens}\right]+\kappa_{\rm 2at}\mathcal{L}\left(\hat{s}^{2}\right)\hat{\rho}_{\rm ens}.
\end{equation}

\begin{figure}[b]
	\centering
	\includegraphics[width=8.3cm]{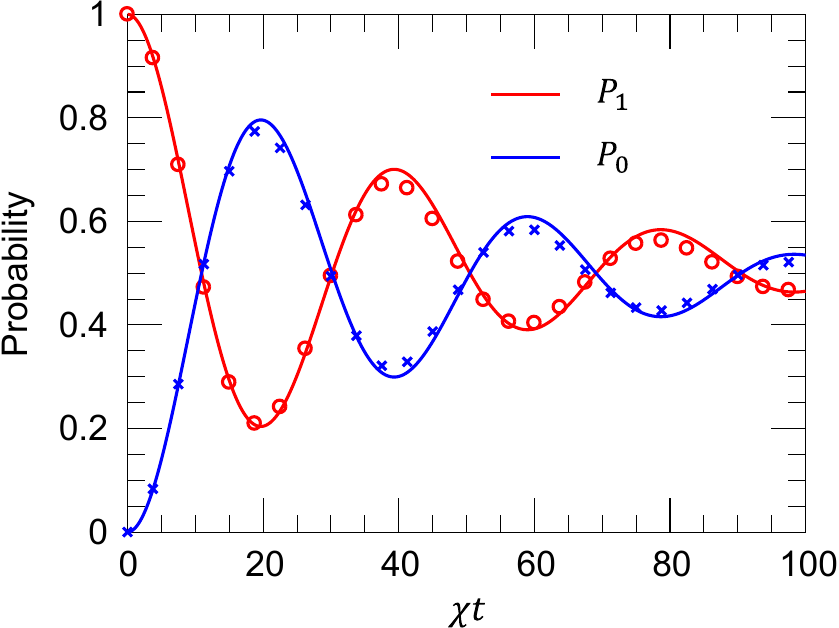}
	\caption{Probabilities, $P_{0}$ and $P_{1}$, of the ensemble being in the states $\ket{0}$ and $\ket{1}$, respectively, for $\Omega_{d}=0.1\kappa_{\rm 2at}$ and $\theta_{d}=0$. The Rabi oscillations between the states $\ket{0}$ and $\ket{1}$ are demonstrated. Symbols are the exact results obtained by integrating the original master equation in Eq.~(\ref{eq_full_master_equation}), with the driving Hamiltonian $\hat{\mathcal{H}}_{d}$ in Eq.~(\ref{eq_driving_Hamiltonian}) transformed to the proper frame. Curves are the effective predictions given by the master equation in Eq.~(\ref{eq_adiabatic_master_equation}). The parameters used here are the same as those in Fig.~\ref{fig_fidelity_2D_manifold}.}\label{fig_rabi_oscillation}
\end{figure}

Clearly, the driving $\Omega_{d}$ can cause transitions out of the desired subspace $\mathcal{C}$. However, as long as $\Omega_{d}\ll \kappa_{\rm 2at}$, the two-atom decay can bring the excited out-of-subspace states quickly back to the subspace $\mathcal{C}$ and, as a result, the dominant dynamics of the ensemble is still restricted into $\mathcal{C}$. To see this more clearly, we express the driving Hamiltonian $\hat{\mathcal{H}}_{d}$ and the two-atom decay operator $\hat{s}^{2}$ in terms of the states $\ket{0}$, $\ket{1}$, and $\ket{2}$ as follows:
\begin{align}
\hat{\mathcal{H}}_{d}=\;&\Omega_{d}\left(e^{i\theta_{d}}\ket{1}\bra{0}+e^{i\theta_{d}}\sqrt{2}\ket{2}\bra{1}+{\rm H.c.}\right),\\
\hat{s}^{2}=\;&\sqrt{2}\ket{0}\bra{2}.
\end{align} 
It can be seen that the driving $\Omega_d$ drives the transitions $\ket{0}\leftrightarrow\ket{1}\leftrightarrow\ket{2}$,
but then the state $\ket{2}$ decays to the state $\ket{0}$ via the two-atom decay. This  results in a coherent coupling between the states $\ket{0}$ and $\ket{1}$, and at the same time an effective decay from the state $\ket{1}$ to the state $\ket{0}$. To further quantify these processes, we now adiabatically eliminate the state $\ket{2}$ (see Appendix~\ref{Adiabatic elimination of the state}), and obtain the Hamiltonian 
\begin{equation}\label{eq_adiabatic_Hamiltonian}
\hat{\mathcal{H}}^{\rm qubit}_{d}=\Omega_{d}\left(e^{-i\theta_{d}}\hat{\sigma}_{-}+e^{i\theta_{d}}\hat{\sigma}_{+}\right),
\end{equation}
with the corresponding master equation,
\begin{align}\label{eq_adiabatic_master_equation}
\dot{\hat{\rho}}_{\rm ens}=-i\left[\hat{\mathcal{H}}^{\rm qubit}_{d},\hat{\rho}_{\rm ens}\right]+\gamma\mathcal{L}\left(\hat{\sigma}_{-}\right)\hat{\rho}_{\rm ens},
\end{align}
where $\gamma=4\Omega^{2}_{d}/\kappa_{\rm 2at}$ is the rate of the effective decay of the ensemble qubit. Here, we have defined $\hat{\sigma}_{-}=\ket{0}\bra{1}$ and $\hat{\sigma}_{+}=\hat{\sigma}_{-}^{\dagger}$ as the lowering and raising operators of the ensemble qubit, respectively. As expected, $\hat{\mathcal{H}}^{\rm qubit}_{d}$ represents a coherent driving of the ensemble qubit, which can arbitrarily rotate the state of this ensemble qubit  on the Bloch sphere. 

In Fig.~\ref{fig_rabi_oscillation}, we plot the time evolution of the probabilities, $P_{0}$ and $P_{1}$, of finding the ensemble qubit being in the states $\ket{0}$ and $\ket{1}$, respectively. We find that the exact results are in excellent agreement with the effective predictions. The Rabi oscillations between the states $\ket{0}$ and $\ket{1}$ are clear, implying that the driving $\Omega_{d}$ can effectively manipulate the ensemble qubit. Furthermore, the amplitude of the Rabi oscillations decreases mostly due to the ensemble decay $\gamma$ as induced by the two-atom decay. Note that here the Rabi oscillations are driven by an external driving applied to the atoms, rather than to the pump mode of the DPA.

\section{Inhomogeneous broadening}
\label{Inhomogeneous broadening}
So far, we have considered a model, where there is no spin dephasing. Now, we consider inhomogeneous broadening, which is the main source of noise in typical atomic ensembles. The thermal noise can safely be neglected here, since the number of thermal photons is very low. For example, for NV spin ensembles with an atomic transition frequency of $\omega_q\simeq2\pi\times3$~GHz, the thermal photon number $n_{\rm th}$ is found to be $\simeq0.03$ at a temperature $T=40$~mK, according to the expression $n_{\rm th}=\left[\exp(\hbar \omega_q/k_{B}T)-1\right]^{-1}$.

Inhomogeneous broadening can be described by the following Hamiltonian
\begin{equation}
	\hat{H}_{\rm inh}=\frac{1}{2}\sum_{j=1}^{N}\delta_{j}\hat{\sigma}^{z}_{j},
\end{equation}
where $\delta_{j}=\omega_{j}-\omega_{q}$. Here, $\omega_{j}$ is the transition frequency of the $j$th atom, and $\omega_{q}$ can be considered as the average of transition frequencies of all the atoms, i.e., $\omega_{q}=\sum_{j=1}^{N}\omega_{j}/N$. The detrimental effects of spin dephasing due to inhomogeneous broadening can, in principle, be completely canceled using spin echo pulses~\cite{hahn1950spin}. Below, we show that our proposal can inherently suppress these detrimental effects, without applying spin echo techniques.

\begin{figure}[b]
	\centering
	\includegraphics[width=16cm]{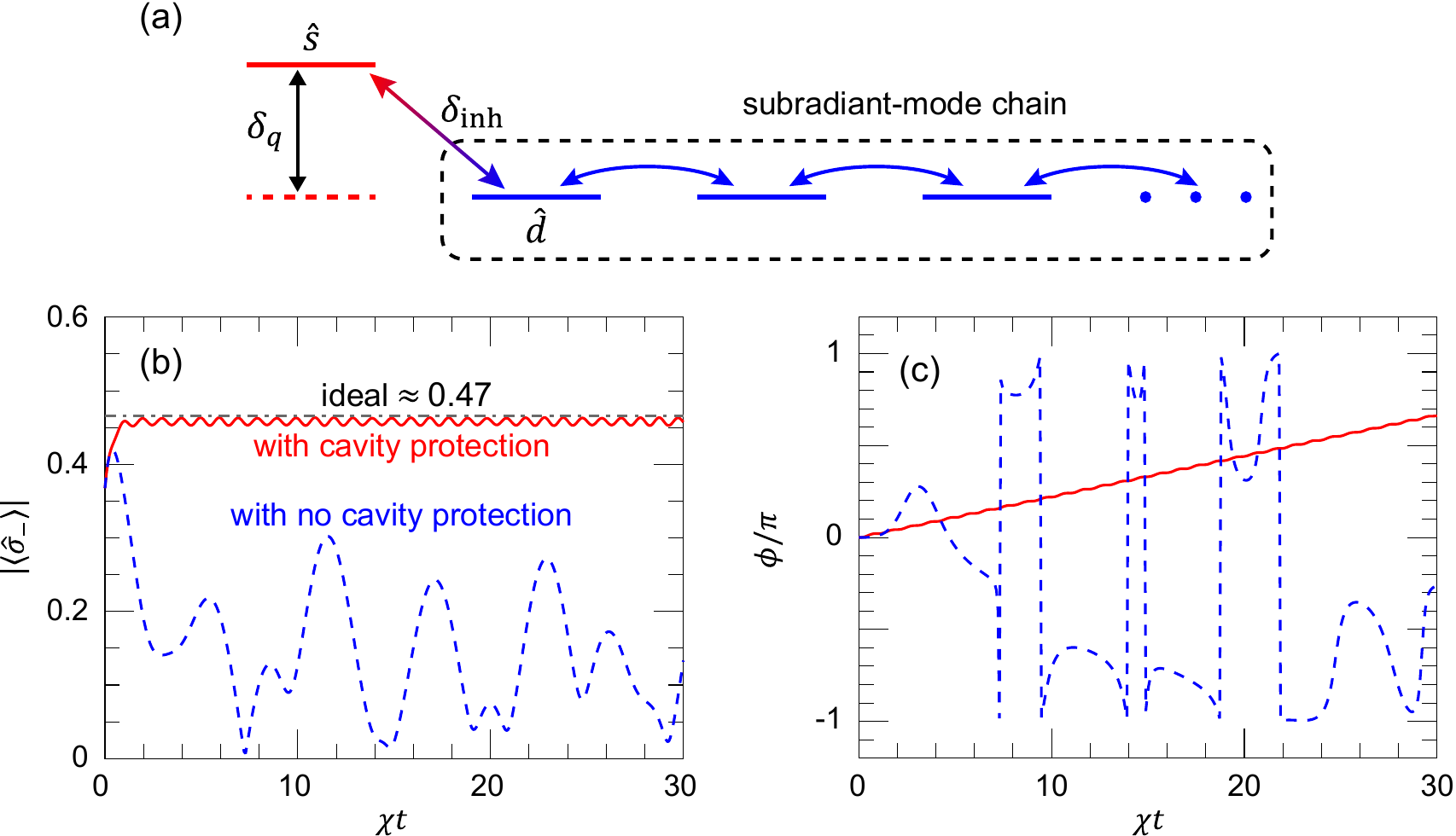}
	\caption{(a) Schematic illustration of suppressing spin dephasing induced by inhomogeneous broadening. The superradiant mode $\hat{s}$ is coupled to the end mode $\hat{d}$ of a chain of subradiant modes, with a strength $\delta_{\rm inh}$ and a detuning $\delta_{q}$. The effect of inhomogeneous broadening can be strongly suppressed, as long as $\delta_{\rm inh}\ll\delta_{q}$. (b), (c) Time evolution of the coherence, $\average{\hat{\sigma}_{-}}=\left|\average{\hat{\sigma}_{-}}\right|\exp(i\phi)$, of the ensemble qubit, with and without the cavity protection. The modulus $\left|\average{\hat{\sigma}_{-}}\right|$ is plotted in (b), and the phase $\phi$ is in (c). Here, we integrated the adiabatic master equation in Eq.~(\ref{eq_two_atom_decay}), with an additional inhomogeneous broadening, and the initial state is assumed to be a spin coherent state of amplitude $\alpha=1$. The horizontal dash-dotted line refers to the ideal value of $\left|\average{\hat{\sigma}_{-}}\right|$, given by Eq.~(\ref{eq_co1_spin_coherent}). We set $N=6$, $\delta_{\rm inh}=0.1\delta_{q}$, and the other parameters are the same as those in Fig.~\ref{fig_fidelity_2D_manifold}.}\label{fig_inhomogeneous_broadening}
\end{figure}

In the low excitation regime, each two-level atom can be modelled by a bosonic mode $b_{j}$~\cite{kurucz2009qubit,kurucz2011spectroscopic,diniz2011strongly,kubo2012storage}, such that
\begin{align}
\hat{\sigma}_{-}^{j}\simeq\;&\hat{b}_{j}, \\ \hat{\sigma}_{z}^{j}\simeq\;&-1+2\hat{b}^{\dagger}_{j}\hat{b}_{j}.
\end{align}
Correspondingly, the Hamiltonian $\hat{H}_{\rm inh}$ becomes
\begin{equation}\label{eq_inhomogeneous_broadening}
\hat{\mathcal{H}}_{\rm inh}=\sum_{j=1}^{N}\delta_{j}\hat{b}^{\dagger}_{j}\hat{b}_{j}.
\end{equation}
According to the definition of the mode $\hat{s}$ given in Eq.~(\ref{eq_superradiant_spin_wave_mode}), it is clearly seen that
\begin{equation}
\hat{s}=\frac{1}{\sqrt{N}}\sum_{j=1}^{N}\hat{b}_{j}.
\end{equation}
Since the mode $\hat{s}$ is coupled to the cavity photon, it is thus called the superradiant spin-wave mode. In addition to this mode, there also exist $(N-1)$ orthogonal dark modes, which are uncoupled from the cavity photon. These dark modes are, thus, called subradiant. In the case of no inhomogeneous broadening, i.e., $\delta_{j}=0$, the superradiant mode $\hat{s}$ is decoupled from the subradiant modes, and it is not affected by them. 

In the presence of inhomogeneous broadening, the Hamiltonian $\hat{\mathcal{H}}_{\rm inh}$ can be modeled as a chain of coupled subradiant modes~\cite{kurucz2011spectroscopic}, as shown in Fig.~\ref{fig_inhomogeneous_broadening}(a), and the superradiant mode $\hat{s}$ is coupled to the subradiant mode 
\begin{equation}
\hat{d}=\frac{1}{\sqrt{N}\delta_{\rm inh}}\sum_{j=1}^{N}\delta_{j}\hat{b}_{j}
\end{equation}
at the end of this chain. Here, $\delta_{\rm inh}$ is the variance of the atomic transition frequencies, and is defined as
\begin{equation}
\delta_{\rm inh}^{2}=\frac{1}{N}\sum_{j=1}^{N}\delta_{j}^{2}.
\end{equation}
Typically, the number $N$ can reach the order of $10^{12}$ as mentioned above, and in this case, the distribution of $\delta_{j}$ is usually considered as continuous, e.g., as a Gaussian or even Poissonian distribution. As a consequence, the variance $\delta_{\rm inh}$ can be considered as a constant here. The coupling strength between the modes $\hat{s}$ and $\hat{d}$ is found to be $\delta_{\rm inh}$, according to the Heisenberg equation of motion for the mode $\hat{s}$. However, as discussed above and in Appendix~\ref{Derivation of the effective Hamiltonian}, the resonance of the superradiant mode $\hat{s}$ is shifted by an amount $\delta_{q}$, due to its detuned coupling to the cavity mode $\hat{a}_{s}$, which is not mediated by virtual photon pairs of the mode $\hat{a}_{s}$. Under the assumption that 
\begin{equation}
\delta_{\rm inh}\ll\delta_{q},
\end{equation}
the coupling between the superradiant mode $\hat{s}$ and the subradiant mode $\hat{b}$ becomes far detuned and, therefore, can be neglected. This is the so-called cavity protection effect~\cite{kurucz2009qubit,kurucz2011spectroscopic,diniz2011strongly,putz2014protecting}. 

In Figs.~\ref{fig_inhomogeneous_broadening}(b) and~\ref{fig_inhomogeneous_broadening}(c), we perform numerical simulations, with and without the cavity protection. Specifically, we integrate the adiabatic master equation in Eq.~(\ref{eq_two_atom_decay}), with the inhomogeneous-broadening Hamiltonian $\hat{\mathcal{H}}_{\rm inh}$ given in Eq.~(\ref{eq_inhomogeneous_broadening}), and then calculate the complex coherence, $\average{\hat{\sigma}_{-}}=\left|\average{\hat{\sigma}_{-}}\right|\exp(i\phi)$, of the ensemble qubit. Note that, here, $\hat{\mathcal{H}}_{\rm inh}$ needs to be transformed, by a unitary operator $\exp\left(i\delta_{q}\hat{s}^{\dagger}\hat{s}t\right)$, to the rotating frame of the adiabatic master equation. In fact, this transformation reflects the suppression effect of the detuning $\delta_{q}$ on the inhomogeneous-broadening induced spin dephasing. The frequency shifts $\delta_{j}$ are randomly chosen according to a Gaussian distribution of mean $0$ and variance $\delta_{\rm inh}$. In Fig.~\ref{fig_inhomogeneous_broadening}(b), we see that the cavity protection stabilizes the modulus $\left|\average{\hat{\sigma}_{-}}\right|$ at almost the ideal value given in Eq.~(\ref{eq_co1_spin_coherent}). Furthermore, Fig.~\ref{fig_inhomogeneous_broadening}(c) shows that, with cavity protection, the phase $\phi$ increases linearly with the evolution time, and thus can be subtracted in a proper frame. In sharp contrast, the coherence of the ensemble qubit (or quantum information) is easily destroyed in the case of no cavity protection. This indicates that our proposal can inherently and strongly suppress the effect of inhomogeneous broadening, as long as $\delta_{\rm inh}\ll\delta_{q}$.

\section{Possible implementations}
We here consider a possible experimental realization utilizing a hybrid system of superconducting circuits and spin ensembles of nitrogen-vacancy (NV) centers. In superconducting circuits, it has been experimentally shown that the available single-photon parametric coupling $J$ can range from hundreds of kHz to tens of MHz~\cite{leghtas2015confining,touzard2018coherent,lescanne2020exponential,chang2020observation,vrajitoarea2020quantum}. For example, in Ref.~\cite{vrajitoarea2020quantum}, two superconducting microwave oscillators are inductively coupled through a radio-frequency superconducting quantum interference device (rf-SQUID), and in that case the value of $J$ can reach $\simeq2\pi\times17.7$~MHz. This setup, with such a strong single-photon parametric coupling, can be used as a fully quantum DPA. Furthermore, by placing an NV spin ensemble, of size $\simeq4.5\times2.25\times0.5$~mm$^{3}$~\cite{amsuss2011cavity,astner2017coherent}, on top of a superconducting microwave oscillator, their strong collective coupling has also been widely demonstrated in experiments; see, e.g., Refs.~\cite{kubo2010strong,amsuss2011cavity,kubo2012storage,putz2014protecting,astner2017coherent}. A typical value of such a collective coupling strength is $g_{\rm col}=2\pi\times10$~MHz, which can be used for our proposal. Hence, the key elements required in our proposal have already been experimentally demonstrated, and we can expect that our proposal can be implemented in a hybrid system of superconducting circuits and NV spin ensembles.

\section{Conclusions}
\label{Conclusions}
We have introduced a method of how to generate an ensemble qubit. The method does not rely on Rydberg dipole blockade, and eliminates the need for the strong atom-atom interactions. The anharmonicity required for the formation of a qubit is provided by the nonlinear two-atom decay. We have demonstrated that the two-atom decay can restrict the ensemble states into a 2D quantum manifold, spanned by the ground state and a symmetric single-excitation superradiant state. The transitions to higher excited superradiant states are strongly suppressed by the two-atom decay, with the residual transitions being converted into an effective decay between the single-excitation superradiant state and the ground state. We have also shown that spin dephasing induced by inhomogeneous broadening can be strongly suppressed, due to the cavity protection effect, with which our proposal is compatible. This indicates that our ensemble qubit can have a long lifetime. We expect that our ensemble qubit could find and stimulate diverse applications in modern quantum technologies.

\section{acknowledgments}
W.Q. was supported in part by the Incentive Research Project of RIKEN. A.M. was supported by the Polish National Science Centre (NCN) under the
Maestro Grant No. DEC-2019/34/A/ST2/00081.
F.N. is supported in part by: 
Nippon Telegraph and Telephone Corporation (NTT) Research, 
the Japan Science and Technology Agency (JST) [via 
the Quantum Leap Flagship Program (Q-LEAP), 
and the Moonshot R\&D Grant Number JPMJMS2061],
the Asian Office of Aerospace Research and Development (AOARD) (via Grant No. FA2386-20-1-4069), and 
the office of Naval Research (ONR).

\appendix 

\setcounter{equation}{0} \setcounter{figure}{0}
\setcounter{table}{0} \makeatletter
\renewcommand{\thefigure}{A\arabic{figure}}

\section{Collective spin operators for nonuniform single-atom couplings}
\label{Collective spin operators for nonuniform single-atom couplings}
For an atomic ensemble with a finite size, the couplings of the atoms to a cavity mode are nonuniform, due to the nonuniform distribution of the electromagnetic field inside the cavity. For such a case, we define the collective spin operators in this Appendix. We now assume that the coupling of the $j$th atom to the cavity mode is $g_{j}$, and the interaction between the ensemble and the cavity can thus be described by the Hamiltonian,
\begin{equation}\label{eq:Hint}
	\hat{H}_{\rm int}=\sum_{j=1}^{N}g_{j}\left(\hat{a}_{s}^{\dagger}\hat{\sigma}_{j}^{-}+\hat{a}_{s}\hat{\sigma}_{j}^{+}\right).
\end{equation}
In this case, we can define the collective spin operators,
\begin{equation}\label{eq:collectiveoperators}
	\hat{S}_{z}=\frac{1}{2}\sum_{j=1}^{N}\hat{\sigma}_{j}^{z}, \quad  {\rm and} \quad \hat{S}_{\pm}=\frac{1}{g}\sum_{j=1}^{N}g_{j}\hat{\sigma}_{j}^{\pm},
\end{equation}
where
\begin{equation}
	g=\sqrt{\frac{1}{N}\sum_{j=1}^{N}g_{j}^{2}}.
\end{equation}
Note that in the main text, the couplings of the atoms to the cavity have been assumed to be uniform for simplicity; that is, $g_{j}=g$.

With the collective operators in Eq.~(\ref{eq:collectiveoperators}), the Hamiltonian $\hat{H}_{\rm int}$ in Eq.~(\ref{eq:Hint}) is accordingly transformed into
\begin{equation}
	\hat{H}_{\rm int}=g\left(\hat{a}_{s}\hat{S}_{+}+\hat{a}_{s}^{\dagger}\hat{S}_{-}\right),
\end{equation}
which, clearly, has the same form as in the uniform case of $g_{j}=g$ [see Eq.~(\ref{eq_full_Hamiltonian})].

\section{Derivation of the time-averaged Hamiltonian in Eq.~(\ref{H_avg})}
\label{Derivation of the effective Hamiltonian}
In this Appendix, we show in detail how to derive the time-averaged Hamiltonian $\hat{\mathcal{H}}_{\rm avg}$. To begin, we rewrite the Hamiltonian in Eq.~(\ref{eq_full_Hamiltonian_boson}) in the interaction picture, yielding
\begin{align}
\hat{\mathcal{H}}_{\rm I}=g_{\rm col}\left(\hat{a}_{s}^{\dagger}\hat{s}e^{i\Delta_{q}t}+\hat{a}_{s}\hat{s}^{\dagger}e^{-i\Delta_{q}t}\right)+J\left(\hat{a}_{p}\hat{a}_{s}^{\dagger 2}e^{i\Delta_{p}t}+\hat{a}_{p}^{\dagger}\hat{a}_{s}^{2}e^{-i\Delta_{p}t}\right),
\end{align}
where $\Delta_{q}=\omega_{s}-\omega_{q}$ and $\Delta_{p}=2\omega_{s}-\omega_{p}$.
According to the formalism of Refs.~\cite{gamel2010time,shao2017generalized},
the second-order and third-order time-averaged Hamiltonians are given, respectively, by
\begin{align}\label{eq_2nd_order_time_averaged_H}
\hat{\mathcal{H}}_{\rm TA}^{\left(2\right)}=\frac{g_{\rm col}^{2}}{\Delta_{q}}\left(\hat{a}_{s}^{\dagger}\hat{a}_{s}-\hat{s}^{\dagger}\hat{s}\right)-\frac{J^{2}}{\Delta_{p}}\left(2\hat{a}_{p}^{\dagger}\hat{a}_{p}+4\hat{a}_{p}^{\dagger}\hat{a}_{p}\hat{a}_{s}^{\dagger}a_{s}-\hat{a}_{s}^{\dagger}\hat{a}_{s}^{\dagger}\hat{a}_{s}a_{s}\right).
\end{align}
and
\begin{equation}
\hat{\mathcal{H}}_{\rm TA}^{\left(3\right)}=\chi\left(\hat{a}_{p}\hat{s}^{\dagger 2}e^{-i\delta t}+\hat{a}_{p}^{\dagger}s^{2}e^{i\delta t}\right).
\end{equation}
Here, $\delta=\omega_{p}-2\omega_{q}$ and $\chi=g_{\rm col}^{2}J/\Delta_{q}^{2}$. Note that as demonstrated below, the detuning $\delta$ is small but nonzero, so as to compensate some undesired shifts induced by the second-order process.

Since the signal mode $\hat{a}_{s}$ is initialized in the vacuum state, the second-order time-averaged Hamiltonian is reduced to
\begin{equation}\label{eq_appendix_resonance_shift}
\hat{\mathcal{H}}_{\rm TA}^{\left(2\right)}=-\frac{g_{\rm col}^{2}}{\Delta_{q}}\hat{s}^{\dagger}\hat{s}
-\frac{2J^{2}}{\Delta_{p}}\hat{a}_{p}^{\dagger}\hat{a}_{p}.
\end{equation}
It is seen that the second-order process induces a Lamb shift~\cite{scully1997book} for atoms (i.e., the first term), and a resonance shift for the pump mode (i.e., the second term). In order to compensate these two undesired shifts, we set 
\begin{equation}
\delta=\frac{2J^{2}}{\Delta_{p}}-\frac{2g_{\rm col}^{2}}{\Delta_{q}}.
\end{equation}
Under this condition, the time-averaged Hamiltonian $\hat{\mathcal{H}}_{\rm avg}$ can be achieved in a proper frame.

\section{Adiabatic elimination of the pump mode $\hat{a}_{p}$}
\label{Adiabatic elimination of the pump mode}
In this appendix, we demonstrate how to adiabatically eliminate the pump mode $\hat{a}_{p}$ of the time-averaged Hamiltonian $\hat{\mathcal{H}}_{\rm avg}$ in Eq.~(\ref{H_avg}). We now consider the following master equation,
\begin{equation}\label{eq_time_averaged_master_equation}
\dot{\hat{\rho}}=-i\left[\hat{\mathcal{H}}_{\rm avg},\hat{\rho}\right]+\kappa_{p}\mathcal{L}\left(\hat{a}_{p}\right)\hat{\rho}.
\end{equation}
Under the assumption of $\kappa_{p}\gg\chi$, the photon population of the pump mode is very low. This allows us to only consider the vacuum state $\ket{0}$ and single-photon state $\ket{1}$ of the pump mode. Here, the number in the ket refers to the number of photons of the pump mode. As a result, the density matrix $\rho$ can be reexpressed as
\begin{equation}
\hat{\rho}=\rho_{00}\ket{0}\bra{0}+\rho_{01}\ket{0}\bra{1}+\rho_{10}\ket{1}\bra{0}+\rho_{11}\ket{1}\bra{1}.
\end{equation}
It follows, upon substituting it into the master equation in Eq.~(\ref{eq_time_averaged_master_equation}), that  
\begin{align}
\dot{\rho}_{00}=\;&i\chi\left(\rho_{01}\hat{s}^{2}-\hat{s}^{\dagger 2}\rho_{10}\right)+\kappa_{p}\rho_{11},\\
\dot{\rho}_{11}=\;&i\chi\left(\rho_{10}\hat{s}^{\dagger 2}-\hat{s}^{2}\rho_{01}\right)-\kappa_{p}\rho_{11},\\
\dot{\rho}_{01}=\;&i\chi\left(\rho_{00}\hat{s}^{\dagger 2}-\hat{s}^{\dagger 2}\rho_{11}\right)-\frac{\kappa_{p}}{2}\rho_{01},
\end{align}
and $\rho_{10}=\rho_{01}^{\dagger}$. Then setting $\dot{\rho}_{01}=0$ gives
\begin{equation}\label{appendix_rho01}
\rho_{01}=i\frac{2}{\kappa_{p}}\chi\left(\rho_{00}\hat{s}^{\dagger 2}-\hat{s}^{\dagger 2}\rho_{11}\right),
\end{equation}
and, in turn, the adiabatic master equation only with the degree of freedom of the ensemble, as follows:
\begin{align}\label{appendix_ensemble_master_equation}
\dot{\hat{\rho}}_{\rm ens}=\dot{\rho}_{00}+\dot{\rho}_{11}=\kappa_{\rm 2at}\mathcal{L}\left(\hat{s}^{2}\right)\hat{\rho}_{\rm ens},
\end{align}
where $\kappa_{\rm 2at}=4\chi^{2}/\kappa_{p}$. Note that in Eq.~(\ref{appendix_ensemble_master_equation}), we have set $\rho_{11}=0$, due to the fact that the photon population of the pump mode is very low.

\section{Adiabatic elimination of the superradiant state $\ket{2}$}
\label{Adiabatic elimination of the state}

We begin with the master equation in Eq.~(\ref{eq_driving_two_atom_decay_master_equation}). In the limit $\Omega_{d}\ll\kappa_{\rm 2at}$, the coupling of the superradiant states $\ket{1}$ and $\ket{2}$, which is induced by the driving $\Omega_{d}$, is considered as only a perturbation. Therefore, the state $\ket{2}$ can be adiabatically eliminated. To do so, we define, according to Ref.~\cite{reiter2012effective}, a non-Hermitian Hamiltonian
\begin{equation}
\hat{\mathcal{H}}_{\rm NH}=-\frac{i}{2}\kappa_{\rm 2at}\hat{s}^{\dagger 2}\hat{s}^{2}=-i\kappa_{\rm 2at}\ket{2}\bra{2},
\end{equation}
and a perturbative coupling 
\begin{equation}
\hat{\mathcal{V}}=\sqrt{2}\Omega_{d} e^{i\theta_{d}}\ket{2}\bra{1}.
\end{equation}
With them, we can obtain an effective coupling between the states $\ket{0}$ and $\ket{1}$, described by
\begin{align}
\hat{\mathcal{H}}^{\rm qubit}_{d}=\;&-\frac{1}{2}\hat{\mathcal{V}}^{\dagger}\left[\hat{\mathcal{H}}_{\rm NH}^{-1}+\left(\hat{\mathcal{H}}_{\rm NH}^{-1}\right)^{\dagger}\right]\hat{\mathcal{V}}+\Omega_{d}\left(e^{-i\theta_{d}}\hat{\sigma}_{-}+e^{i\theta_{d}}\hat{\sigma}_{+}\right),\\
=\;&\Omega_{d}\left(e^{-i\theta_{d}}\hat{\sigma}_{-}+e^{i\theta_{d}}\hat{\sigma}_{+}\right).
\end{align}
and an effective Lindblad operator 
\begin{equation}
2\kappa_{\rm 2at}\mathcal{L}\left(\ket{0}\bra{2}\hat{\mathcal{H}}_{\rm NH}^{-1}\hat{\mathcal{V}}\right)=\gamma\mathcal{L}\left(\hat{\sigma}_{-}\right),
\end{equation}
where $\gamma=4\Omega^{2}_{d}/\kappa_{\rm 2at}$.
Thus, the master equation in Eq.~(\ref{eq_adiabatic_master_equation}) can be achieved.


%

\end{document}